\documentclass[11pt]{article} 
\usepackage{hyperref}

\usepackage[utf8]{inputenc} 

\usepackage{geometry} 
\usepackage{graphicx} 

\usepackage[round]{natbib}

\usepackage{booktabs} 
\usepackage{array} 
\usepackage{paralist} 
\usepackage{verbatim} 
\usepackage{subfig} 

\usepackage{amsmath}
\usepackage{amsfonts}
\usepackage{amssymb}
\usepackage{mathrsfs}
\usepackage{bbm}
\usepackage{amsthm}

\usepackage{multirow}

\newtheorem{theorem}{Theorem}[section]
\newtheorem{corollary}[theorem]{Corollary}

\theoremstyle{definition}
\newtheorem{definition}[theorem]{Definition}
\theoremstyle{remark}

\theoremstyle{remark}
\newtheorem{example}[theorem]{Example}
\numberwithin{equation}{section}

\allowdisplaybreaks

\newcommand{\RRR}{\mathbb{R}}
\newcommand{\EEE}{\mathbb{E}}

\newcommand{\dd}{\mathrm{d}}
\newcommand{\M}{M}
\newcommand{\rA}{\mathrm{A}}
\newcommand{\rB}{\mathrm{B}}
\newcommand{\supp}{\mathrm{supp}}
\newcommand{\cF}{\mathcal{F}}
\newcommand{\cN}{\mathcal{N}}

\newcommand{\rT}{\mathrm{T}}

\newcommand{\CRPS}{\mathrm{CRPS}}
\newcommand{\QSF}{\mathrm{QSF}}
\newcommand{\ESF}{\mathrm{ESF}}
\newcommand{\HSF}{\mathrm{HSF}}

\newcommand{\one}{\mathbbm{1}}

\newcommand{\gjrect}{
	g_j(t) = 
	\begin{cases}
	0, & t< a\\
	t-a, & a \leq t < b \\
	 b-a, & t \geq  b
	\end{cases}
}
\newcommand{\gjtrap}{
	g_j(t) = 
	\begin{cases}
	0, & t<a \\
	(t-a)^2/(2(b-a)), & a \leq t < b \\
	t - (b+a)/2, & b\leq t<c\\
	-(d-t)^2/(2(d-c))+(d+c-a-b)/2, & c \leq t < d \\
	(d+c-a-b)/2, & t \geq d
	\end{cases} 
}
\newcommand{\phijrect}{
	\phi_j(t) =
	\begin{cases}
	0, & t<a\\
	2(t-a)^2, & a \leq t < b \\
	4(b-a)t + 2(a^2-b^2), & t \geq b
	\end{cases}
}
\newcommand{\phijtrap}{
	\phi_j(t) = 
	\begin{cases}
	0, & t<a \\
	\frac{2(t-a)^3}{3(b-a)}, & a \leq t < b \\
	2t^2 - 2(a+b)t+\frac23(b-a)^2+2ab, & b\leq t<c\\
	\frac{2(d-t)^3}{3(d-c)}+2(d+c-a-b)t+\tfrac23((b-a)^2+3ab-(d-c)^2-3cd), & c \leq t < d \\
	2(d+c-a-b)t + \tfrac23((b-a)^2+3ab-(d-c)^2-3cd), & t \geq d
	\end{cases}
}
\newcommand{\mat}[2]{
	\begin{matrix}
		\text{#1} \\
		\text{given by} \\
		\text{#2}
	\end{matrix}
}

\usepackage{fancyhdr} 
\usepackage{sectsty}
\allsectionsfont{\sffamily\mdseries\upshape} 

\usepackage[nottoc,notlof,notlot]{tocbibind} 
\usepackage[titles,subfigure]{tocloft} 

\title{Evaluation of point forecasts for extreme events using consistent scoring functions}

\author{Robert J. Taggart\\Bureau of Meteorology\\robert.taggart@bom.gov.au}

\begin{document}

\maketitle

\begin{abstract}
\noindent We present a method for comparing point forecasts in a region of interest, such as the tails or centre of a variable's range. This method cannot be hedged, in contrast to conditionally selecting events to evaluate and then using a scoring function that would have been consistent (or proper) prior to event selection. Our method also gives decompositions of scoring functions that are consistent for the mean or a particular quantile or expectile. Each member of each decomposition is itself a consistent scoring function that emphasises performance over a selected region of the variable's range. The score of each member of the decomposition has a natural interpretation rooted in optimal decision theory. It is the weighted average of economic regret over user decision thresholds, where the weight emphasises those decision thresholds in the corresponding region of interest.

\vspace{10pt}

\end{abstract}

\section{Introduction}

Extreme events occur in many systems, from atmospheric to economic, and present significant challenges to society. Hence the accurate prediction of extreme events is of vital importance. In many such situations, competing forecasts are produced by a variety of forecast systems and it is natural to want to compare the performance of such forecasts with emphasis on the extremes.

In this context, it is critical that the methodology for requesting, evaluating and ranking competing forecasting systems is decision-theoretically coherent. Since the future is not precisely known, ideally forecasts ought to be probabilistic in nature, taking the form of a predictive probability distribution and assessed using a proper scoring rule (\citealt{gneiting2014probabilistic}). Nonetheless, in many contexts and for a variety of reasons, point forecasts (i.e. single-valued forecasts taking values from the prediction space) are issued and used. For decision-theoretically coherent evaluation of point forecasts, either the scoring function (such as the squared error or absolute error scoring function) that will be used to assess predictive performance should be advertised in advance, or a specific functional (such as the mean or median) of the forecaster's predictive distribution should be requested and evaluation then conducted using a scoring function that is consistent for that functional (\citealt{gneiting2011making}). The use of proper scoring rules and consistent scoring functions encourage forecasters to quote honest, carefully considered forecasts.

To compare competing forecasts for the extremes, one may be tempted to use what would otherwise be a proper scoring rule or consistent scoring function, but restrict evaluation to a subset of events for which extremes were either observed, or forecast or perhaps both. However, such methodologies promote hedging strategies (as illustrated in Section~\ref{s:hedging}) and can result in misguided inferences. This gives rise to the \textit{forecaster's dilemma}, whereby ``if forecast evaluation proceeds conditionally on a catastrophic event having been observed [then] always predicting a calamity becomes a worthwhile strategy'' (\citealt{lerch2017forecaster}).

Nonetheless, for predictive distributions \cite{gneiting2011comparing} showed that a suitable alternative exists in the \textit{threshold-weighted continuous ranked probability score}, which is a weighted version of the commonly used continuous ranked probability score (CRPS). The weight is selected to emphasise the region of interest (such as the tails or some other region of a variable's range) and induces a proper scoring rule. This technique extends in a natural way to point forecasts targeting the median functional, since the CRPS is a generalisation of the absolute error scoring function. The UK Met Office has recently applied this method to report the performance of temperature forecasts, with emphasis on climatological extremes (\citealt{sharpe2020new}).

Despite this progress, \cite{lerch2017forecaster} offer this summary of the general situation for evaluating point forecasts at the extremes: ``there is no obvious way to abate the forecaster’s dilemma by adapting existing forecast evaluation methods appropriately, such that particular emphasis can be put on extreme outcomes.'' In this paper we remedy this situation. We construct consistent scoring functions that can be used evaluate point forecasts that emphasise performance in the region of interest for important classes of functionals, including expectations and quantiles. Moreover, the relevant specific case of these constructions agrees with the extension of the threshold-weighted CRPS to the median functional.

The main idea of this paper can be illustrated using the squared error scoring function $S(x,y)=(x-y)^2$, which is consistent for point forecasts of the expectation functional. Suppose that the outcome space $\RRR$ is partitioned as $\RRR =  I_1\cup I_2$, where $I_1= (-\infty,a)$ and $I_2=[a,\infty)$ for some $a$ in $\RRR$. Corollary~\ref{cor:decomposition} gives the decomposition $S=S_1+S_2$, where each scoring function $S_i$ is consistent for expectations whilst also emphasising predictive performance on the interval $I_i$. In particular, if $x,y\in I_1$ then $S_2(x,y)=0$. Given a point forecast $x$ and corresponding observation $y$, the explicit formula for $S_2$ is
\begin{equation}\label{eq:S_2 explicit}
S_2(x,y) = (y-a)^2\one\{y\geq a\} - (x- a)^2\one\{x\geq a\} - 2(y-x)(x-a)\one\{x\geq a\}.
\end{equation}
Here $\one$ denotes the indicator function, so that $\one\{x\geq a\}$ equals $1$ if $x\geq a$ and $0$ otherwise.

\begin{figure}[bt]
\centering
\includegraphics{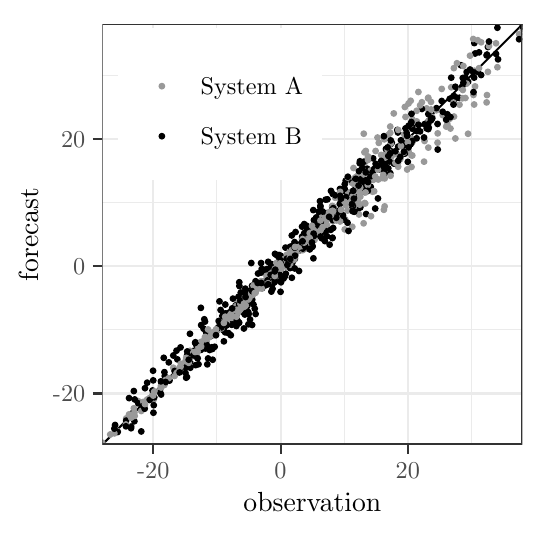}
\caption{Scatter plot of forecasts against observations for a random sample of events from the Synthetic data example of Section~3.5.}
\label{fig:synthetic}
\end{figure}

The performance of competing point forecasts for expectations can then be compared by computing the mean scores $\bar S$, $\bar S_1$ and $\bar S_2$ for each forecast system, over the same set events. To illustrate, consider two forecast systems A and B, whose error characteristics are depicted by the scatter plot of Figure~\ref{fig:synthetic}. System B is homoscedastic (i.e., has even scatter about the diagonal line throughout the variable's range) whereas System A is heteroscedastic  (with relatively small errors over lower parts of the variable's range and relatively large over higher parts). For this set of events, the mean squared error $\bar S$ for each system is very similar and there is no statistical significance in their difference. However, with $a=10$, the mean score $\bar S_2$ of System A is significantly higher than that of B (i.e., B is superior when emphasis is placed on the region $[10,\infty)$) whilst the mean score $\bar S_1$ of System A is significantly lower than that of B (i.e., A is superior when emphasis is placed on the region $(-\infty,10)$). Full details for this case study are given in Section~\ref{ss:examples}.

This example is illustrative. Decompositions of the outcome space can consist of more than two intervals, and the boundary between intervals can also be `blurred' by selecting suitable weight functions. Each decomposition of the outcome space then induces a decomposition
\begin{equation}\label{eq:decomp}
S=\sum_{i=1}^n S_i
\end{equation}
of a given consistent scoring function $S$, whose summands $S_i$ are also consistent for the functional of concern and with each $S_i$ emphasising forecast performance in the relevant region of the outcome space. Such decompositions are presented for the consistent scoring functions of quantiles, expectations, expectiles, and Huber means. The approach is unified, in that the same decomposition of the outcome space induces decompositions of the CRPS and of a variety of scoring functions for point forecasts. Details are given in Section~\ref{s:decomposition} with illustrative examples.

Mathematically, the main result of this paper is a corollary of the mixture representation theorems for the consistent scoring functions of quantiles, expectiles (\citealt{ehm2016quantiles}) and Huber functionals (\citealt{taggart2021point}). This furnishes each member $S_i$ of the decomposition (\ref{eq:decomp}) with an interpretation rooted in optimal decision theory. Certain classes of optimal decision rules elicit a user action if and only if a point forecast $x$ exceeds a particular decision threshold $\theta$. The score $S_i(x,y)$ is a measure of economic regret, relative to decisions based on a perfect forecast, of using the point forecast $x$ when the observation $y$ realises, averaged over all decision thresholds $\theta$ belonging to corresponding interval $I_i$ of the partition. Precise details are given in Section~\ref{s:murphy} and illustrated with the aid of Murphy diagrams.

\section{Hedging strategies for na\"ive assessments of forecasts of extreme events}\label{s:hedging}

We illustrate how a seemingly natural approach for comparing predictive performance at the extremes creates opportunities for hedging strategies.

A meteorological agency maintains two forecast systems A and B, each of which produces predictive distributions for hourly rainfall $Y$ at a particular location. System A is fully automated and hence cheaper to support than B. System B has knowledge of the System A forecast prior to issuing its own forecast. The agency requests the mean value of their predictive distributions with a lead time of 1 day. These point forecasts are assessed using the squared error scoring function, which is consistent for forecasts of the mean (i.e., the forecast strategy that minimises one's expected score is to issue the mean value of one's predictive distribution). For a two year period, the bulk of observed and forecast values fall in the interval $[0\,\mathrm{mm}, 10\,\mathrm{mm}]$ and there is no statistically significant difference between the performance of the two systems when scored using the squared error function. 

However, the maintainers of System B claim that B performs better for the extremes and that bulk verification statistics fail to highlight this. The agency decides to use forecasts from A for the majority of cases, but will test whether B is significantly better than A at forecasting the heavy rainfall events. The agency considers four options for selecting hourly events to assess, after which the squared error scoring function will be used to compare predictive performance on those events. If System B does not perform significantly better than A over the next 12 months then it will be decommissioned. 

For a given forecast case, let $F_\rA$ and $F_\rB$ denote the predictive distributions produced by each system, and $x_\rA$ and $x_\rB$ the respective point forecasts issued. Suppose that the random variable $Y$ has a distribution specified by $F_\rB$. For each option, the maintainers of System B have a strategy to optimise their expected score; that is, to choose $x_\rB$ such that $\EEE[(x_\rB-Y)^2]$ is minimised.

\begin{enumerate}
\item[Option 1:] Only assess events where the observation is at least 20 mm.\\Strategy: If $\mathbb{P}(Y\geq20)>0$ then $F_\rB|\{Y\geq20\}$, which denotes the predictive distribution of $\rB$ conditioned on the event $\{Y\geq20\}$, exists. In this case, forecast $x_\rB=\mathrm{Mean}(F_\rB|\{Y\geq20\}))$ and otherwise forecast $x_\rB=20$.
\item[Option 2:] Only assess events where either $x_\rA$ or $x_\rB$ is at least 20 mm.\\Strategy: If $\max(x_\rA, \mathrm{Mean}(F_\rB))\geq 20$ then $x_\rB=\mathrm{Mean}(F_\rB)$. Otherwise if $\EEE[(20-Y)^2]<\EEE[(x_\rA-Y)^2]$ then $x_\rB=20$ else $x_\rB=\mathrm{Mean}(F_\rB)$. This will ensure that the event is assessed whenever System B expects that a forecast of 20 will receive a more favourable score than than a forecast of $x_\rA$.
\item[Option 3:] Only assess events where either $x_\rA$, $x_\rB$ or the observation is at least 20 mm.\\Strategy: If $\max(x_\rA, \mathrm{Mean}(F_\rB))\geq 20$ then $x_\rB=\mathrm{Mean}(F_\rB)$. Else if $\EEE[(20-Y)^2]<\EEE[(x_\rA-Y)^2]$ then $x_\rB=20$. Otherwise, the only other way the event will be assessed is if the observation is at least 20mm, so forecast $x_\rB=19.9$.
\item[Option 4:]  Only assess events where $x_\rA \geq 20$.\\Strategy: In this case there is nothing that System B can to do influence which events will be assessed. Therefore $x_\rB=\mathrm{Mean}(F_\rB)$.
\end{enumerate}

Of these, only Option 4 does not expose the agency to a hedging strategy from System B. However, under Option 4 the developers of A may employ the strategy of forecasting $x_\rA= \max(19.9, \mathrm{Mean}(F_\rA))$, so that no further comparison of systems is made.

There is an Option 5: use a scoring function that is consistent for the mean functional and that emphasises performance at the extremes. We turn attention to this now.

\section{Decompositions of scoring functions}\label{s:decomposition}

We work in a setting where point forecasts $x$ and observations $y$ belong to some interval $I$ of the real line $\RRR$ (possibly with $I=\RRR$). In Section~\ref{ss:partitions of unity} we introduce \textit{partitions of unity}, which are used to `subdivide' the outcome space $I$ into subregions of interest. We then illustrate in Section~\ref{ss:crps} how such subdivisions induce decompositions of the CRPS, where each member of the decomposition emphasises performance on the corresponding subregion of $I$ whilst retaining propriety. To obtain analogous decompositions for consistent scoring functions, we recall in Section~\ref{ss:consistent scoring functions} that the consistent scoring functions for quantiles and expectations (among others) have general mathematical forms. The aim is to find which specific instances of the general form emphasise performance on the subregions specified by the partition of unity. This is answered in Section~\ref{ss:decomposition of sf}. Section~\ref{ss:examples} contains examples of such decompositions, and opens with a worked example showing how to find the formula for the squared error decomposition of Equation (\ref{eq:S_2 explicit}).

\subsection{Partitions of unity}\label{ss:partitions of unity}

Recall that the support of a function $\chi:I \to\RRR$, denoted $\supp(\chi)$, is defined by 
\[\supp(\chi) = \{t\in I: \chi(t)\neq 0\}.\]
In this paper, we say that $\{\chi_j\}_{j=1}^n$ is a \textit{partition of unity} on $I$ if $\{\chi_j\}_{j=1}^n$ is a finite set of measurable functions $\chi_j:I\to\RRR$ such that $0\leq\chi_j(t)\leq 1$ and
\[\sum_{j=1}^n \chi_j(t) = 1\]
whenever $t\in I$. (For readers unfamiliar with the concept of measurability, any piecewise continuous function is measurable.) We will call each $\chi_j$ a \textit{weight function}. We note that these differ from typical definitions in that we do not require $\chi_j$ to be continuous or have bounded support.

\begin{figure}[bt]
\centering
\includegraphics{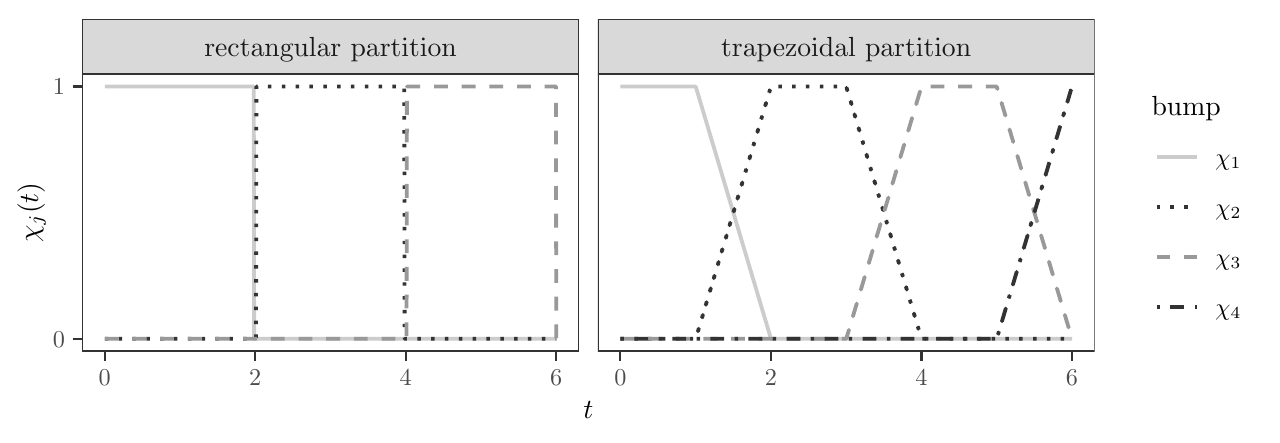}
\caption{Two partitions $\{\chi_j\}_{j=1}^n$ of unity on the interval $[0,6)$, one using rectangular weight functions and the other trapezoidal weight functions.}
\label{fig:partition}
\end{figure}

Figure~\ref{fig:partition} illustrates two different partitions of unity for the interval $I=[0,6)$. The \textit{rectangular partition of unity} is consists of \textit{rectangular weight functions}, each of the form
\begin{equation}\label{eq:rectangular bump}
\chi_j(t) =
\begin{cases}
1, & a\leq t<b\\
0, & \text{otherwise}
\end{cases}
\end{equation}
for suitable constants satisfying $a<b$, both depending on $j$. The \textit{trapezoidal partition of unity} consists of \textit{trapezoidal weight functions}, each typically having the form
\begin{equation}\label{eq:trapezoidal bump}
\chi_j(t) = 
\begin{cases}
(t-a)/(b-a), & a \leq t < b \\
1, & b \leq t < c \\
(d-t)/(d-c), & c \leq t < d \\
0, & \text{otherwise}
\end{cases}
\end{equation}
for suitable constants satisfying $a<b<c<d$, all depending on $j$, with appropriate modification for the end cases.

More generally, if $\{\psi_j\}_{j=1}^n$ is a set of piecewise nonnegative measurable functions with the property that
\[\sum_{j=1}^n\psi_j(t)>0\]
whenever $t\in I$, then a partition of unity $\{\chi_j\}_{j=1}^n$ can be constructed by defining
\[\chi_j(t) = \psi_j(t)\Big(\sum_{j=1}^n\psi_j(t)\Big)^{-1}.\]

\subsection{Decomposition of the CRPS}\label{ss:crps}

Each partition $\{\chi_j\}_{j=1}^n$ of unity induces a corresponding decomposition of the CRPS. Given a predictive distribution $F$, expressed as a cumulative density function, and a corresponding observation $y$, the CRPS is defined by
\[ \CRPS(F,y) = \int_I (F(z) - \one\{y\leq z\})^2\,\dd z\]
and for each $\chi_j$ the threshold-weighted CRPS by
\[\CRPS_j(F,y) = \int_I (F(z) - \one\{y\leq z\})^2\,\chi_j(z)\,\dd z.\]
Both are proper scoring rules (\citealt{gneiting2011comparing}). Thus the $\CRPS$ has a decomposition
\[\CRPS = \sum_{j=1}^n \CRPS_j,\]
where each component $\CRPS_j$ is proper and emphasises performance in the region determined by the weight $\chi_j$. The Sydney rainfall forecasts example of Section~\ref{ss:examples} illustrates the application of such a decomposition. We now establish analogous decompositions for a wide range of scoring functions. 

\subsection{Consistent scoring functions}\label{ss:consistent scoring functions}

For decision-theoretically coherent point forecasting, forecasters need a directive in the form of a statistical functional (\citealt{gneiting2011making}) or a scoring function which should be minimised (\citealt{patton2020comparing}). A \textit{statistical functional} $\rT$ is a (potentially set-valued) mapping from a class of probability distributions $\cF$ to the real line $\RRR$. Examples include the mean (or expectation) functional, quantiles and expectiles (\citealt{newey1987asymmetric}), the latter recently attracting interest in risk management (\citealt{bellini2017risk}). A \textit{consistent scoring function} is a special case of a proper scoring rule in the context of point forecasts, and rewards forecasters who make careful honest forecasts.

\begin{definition} (\citealt{gneiting2011making})
A scoring function $S: I\times I\to[0,\infty)$ is \textit{consistent} for the functional $\rT$ relative to a class $\cF$ of probability distributions if
\begin{equation}\label{eq:consistency}
\EEE S(t,Y) \leq \EEE S(x,Y), \qquad\text{whenever }  Y\sim F,
\end{equation}
for all probability distributions $F\in\cF$, all $t\in \rT(F)$ and all $x\in I$. The function $S$ is \textit{strictly consistent} if $S$ is consistent and if equality in Equation~(\ref{eq:consistency}) implies that $x\in\rT(F)$.
\end{definition}

The consistent scoring functions for many commonly used statistical functionals have general forms.

Given $g:I\to\RRR$ and $\alpha\in(0,1)$, define the `quantile scoring function' $\QSF(g,\alpha):I\times I\to\RRR$ by
\begin{equation}\label{eq:QSF}
\QSF(g,\alpha)(x,y) = (\one\{y<x\} - \alpha)(g(x) - g(y)) \qquad\forall x,y\in I.
\end{equation}
The name QSF is justified because, subject to slight regularity conditions, a scoring function $S$ is consistent for the $\alpha$-quantile functional if and only if $S=\QSF(g,\alpha)$ where $g$ is nondecreasing (\citealt{gneiting2011quantiles, gneiting2011making, thomson1979eliciting}). The  absolute error scoring function $S(x,y)=|x-y|$ for the median functional arises from Equation~(\ref{eq:QSF}) when $g(t)=2t$ and $\alpha=1/2$. The commonly used $\alpha$-quantile scoring function
\begin{equation}\label{eq:standard qsf}
S(x)=(\one\{y<x\}-\alpha)(x-y)
\end{equation}
arises when $g(t)=t$.

Given a convex function $\phi:I\to\RRR$ with subderivative $\phi'$ and $\alpha\in(0,1)$, define the function $\ESF(\phi,\alpha):I\times I\to\RRR$ by
\begin{equation}\label{eq:ESF}
\ESF(\phi,\alpha)(x,y) = |\one\{y<x\} - \alpha|\big(\phi(y) - \phi(x) - \phi'(x)(y-x)\big) \qquad\forall x,y\in I.
\end{equation}
 (The subderivative is a generalisation of the derivative for convex functions and coincides with the derivative when the convex function is differentiable.) Subject to weak regularity conditions, a scoring function $S$ is consistent for the $\alpha$-expectile functional if and only if $S=\ESF(\phi,\alpha)$ where $\phi$ is convex (\citealt{gneiting2011making, savage1971elicitation}). The expectation (or the mean) functional corresponds to the special case $\alpha=1/2$, with the squared error scoring function $S(x,y)=(x-y)^2$ for expectations arising from Equation~(\ref{eq:ESF}) when $\phi(t)=2t^2$ and $\alpha=1/2$. A special case of the squared error scoring function is the Brier score, where $I=[0,1]$ and observations typically take values in $\{0,1\}$.

Given $\phi:I\to\RRR$ with subderivative $\phi'$ and $\nu>0$, define the function $\HSF(\phi,\nu):I\times I\to\RRR$ by
\begin{equation}\label{eq:HSF}
\HSF(\phi,\nu)(x,y) =\tfrac12\big(\phi(y) - \phi(\kappa_\nu(x-y)+y) + \kappa_\nu(x-y)\phi'(x)\big) \qquad\forall x,y\in I,
\end{equation}
where $\kappa_\nu$ is the `capping' function defined by $\kappa_\nu(x)=\max(-\nu,\min(x,\nu))$. Subject to slight regularity conditions, a scoring function $S$ is consistent for the Huber mean functional (with tuning parameter $\nu$) if and only if $S=\HSF(\phi,\nu)$ where $\phi$ is convex (\citealt{taggart2021point}). The Huber loss scoring function
\[
S(x,y)=
\begin{cases}
\tfrac12(x-y)^2, & |x-y|\leq\nu \\
\nu|x-y|-\tfrac12\nu^2, & |x-y|>\nu
\end{cases}
\]
arises from $\HSF(\phi,\nu)$ when $\phi(t)=t^2$, and is used by the Bureau of Meteorology to score forecasts of various parameters. The Huber mean is an intermediary between the median and the mean functionals, and is a robust measure of the centre of a distribution (\citealt{huber1964robust, taggart2021point}).

\subsection{Decomposition of consistent scoring functions}\label{ss:decomposition of sf}

We now state the main result of this paper, namely that the consistent scoring functions for the quantile, expectile and Huber mean functionals can be written as a sum of consistent scoring functions with respect to the chosen partition of unity. It is presented as a corollary since it follows from the mixture representation theorems of \cite{ehm2016quantiles} and \cite{taggart2021point}.

\begin{corollary}\label{cor:decomposition}
Suppose that $\{\chi_j\}_{j=1}^n$ is a partition of unity on $I$. For each $j$ in $\{1,\ldots,n\}$, fix any points $u_j$ and $v_j$ in $I$.
\begin{enumerate}
\item[(a)] If $g:I\to\RRR$ is a nondecreasing differentiable function and $\alpha\in(0,1)$ then
\[\QSF(g,\alpha) = \sum_{j=1}^n \QSF(g_j,\alpha)\]
where $g_j$ is nondecreasing and defined by
\begin{equation}\label{eq:g_j}
g_j(u) = \int_{u_j}^u \chi_j(\theta)g'(\theta)\,\dd\theta.
\end{equation}
Moreover, if $I_0\subset I$ is an interval and $\supp(\chi_j) \cap I_0 = \emptyset$ then $\QSF(g_j,\alpha)=0$ on $I_0\times I_0$.
\item[(b)] If $\phi:I\to\RRR$ is a convex twice-differentiable function, $\alpha\in(0,1)$ and $\nu>0$ then
\begin{equation}\label{eq: ESF decomposition}
\ESF(\phi,\alpha) = \sum_{j=1}^n \ESF(\phi_j,\alpha)
\end{equation}
and
\[\HSF(\phi,\nu) = \sum_{j=1}^n \HSF(\phi_j,\nu),\]
where $\phi_j$ is convex and defined by
\begin{equation}\label{eq:phi_j general}
\phi_j(u) = \int_{u_j}^u \int_{v_j}^v \chi_j(\theta)\phi''(\theta)\,\dd\theta\,\dd v.
\end{equation}
Moreover, if $I_0\subset I$ is an interval and $\supp(\chi_j) \cap I_0 = \emptyset$ then $\ESF(\phi_j,\alpha)=\HSF(\phi_j,\alpha)=0$ on $I_0\times I_0$.
\end{enumerate}
\end{corollary}

See Appendix~\ref{s:proof} for the proof. Appendix~\ref{s:g_j and phi_j} states closed-form expressions for $g_j$ and $\phi_j$ for commonly used scoring functions when the partition of unity is rectangular or trapezoidal. We note that the natural analogue of Corollary~\ref{cor:decomposition} for the consistent scoring functions of Huber functionals (which are generalised Huber means) also holds.

One can show that $\QSF(g_j,\alpha)$, $\ESF(\phi_j,\alpha)$ and $\HSF(\phi_j,\nu)$ are independent of the choice of points $u_j$ and $v_j$ in $I$. In practice, the choice of $u_j$ and $v_j$ may be determined for computational convenience, such as selecting (if it exists) the minimum of $\supp(\chi_j)$. 

Finally, we discuss strict consistency. Suppose that $\QSF(g,\alpha)$, $\ESF(\phi,\alpha)$ and $\HSF(\phi,\nu)$ are strictly consistent for the quantile, expectile or Huber mean functionals respectively for some class $\cF$ of probability distributions. This occurs when $g$ is strictly positive, or when $\phi$ is strictly convex, perhaps subject to mild regularity conditions on $\cF$ (\citealt{gneiting2011making, taggart2021point}). If $\chi_j$ is strictly positive on $I$ then $\QSF(g_j,\alpha)$, $\ESF(\phi_j,\alpha)$ and $\HSF(\phi_j,\nu)$ are also strictly consistent for their respective functionals. An example of such a partition $\{\chi_1,\chi_2\}$ of unity on $\RRR$ is given by
\[\chi_2(t) = \tfrac12 + \tfrac1\pi \arctan(t-a)\]
and $\chi_1(t)=1-\chi_2(t)$. Here, $\chi_2$ induces scoring functions $S_2$ that emphasise performance on $(a,\infty)$ but do not completely ignore performance on any subinterval of $\RRR$.

\subsection{Examples}\label{ss:examples}

We begin by demonstrating how to find explicit formulae for a particular decomposition.

\begin{figure}[bt]
\centering
\includegraphics{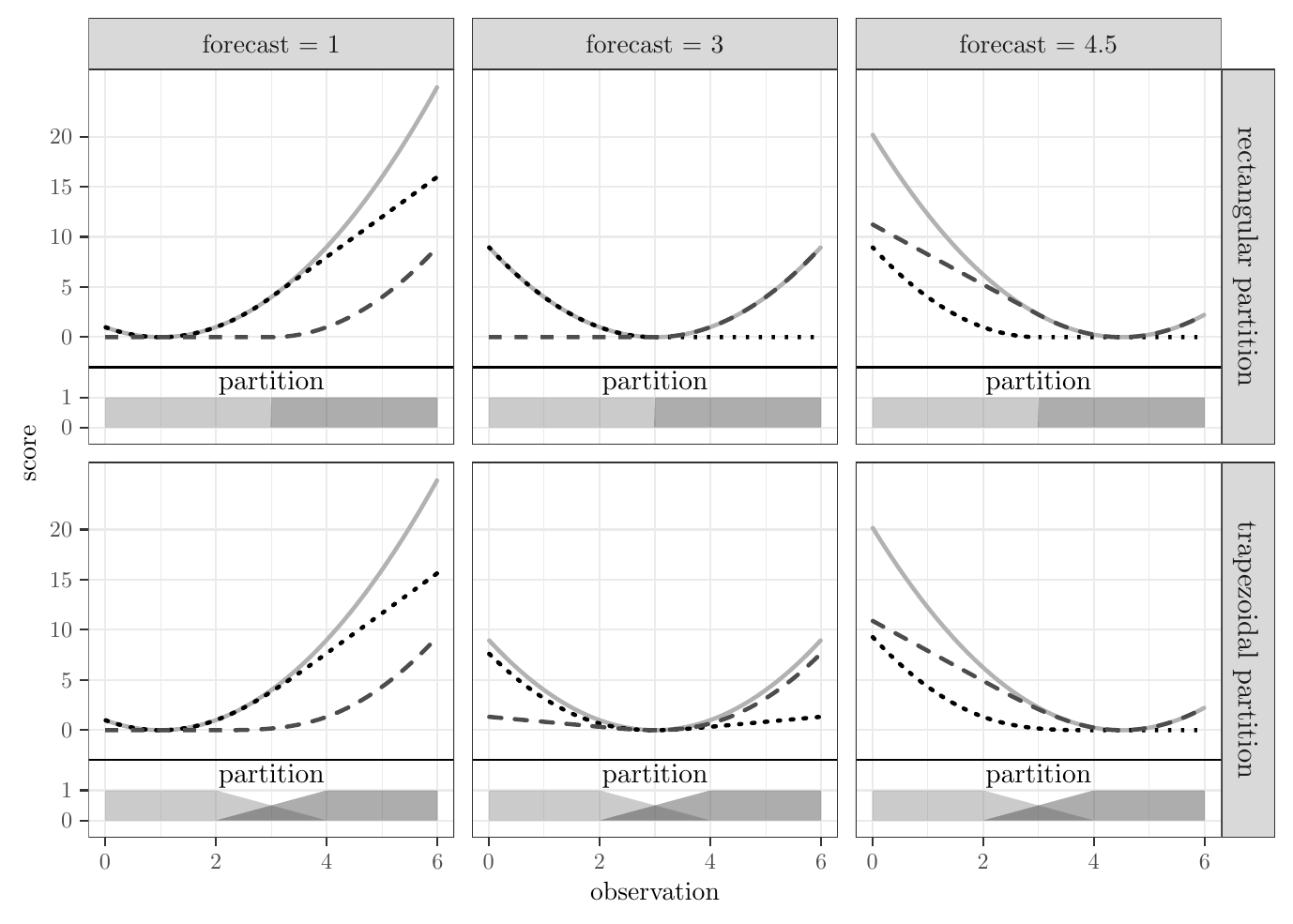}
\caption{Decomposition $S=S_1+S_2$ of the squared error scoring function, using the rectangular and trapezoidal partitions. The solid line represents $S$, the dotted line $S_1$ and the dashed line $S_2$ for different forecast--observation pairs. In each case, $S_1$ emphasises performance on the interval $(0,3)$ while $S_2$ emphasises performance on $(3,6)$.}
\label{fig:esf decomp}
\end{figure}

\begin{example}[Decomposition of the squared error scoring function]\label{ex:esf}
Let $S$ denote the scoring function $S(x,y) = (x-y)^2$. Note that $S=\ESF(\phi,0.5)$, where  $\phi(t) = 2t^2$ via Equation~(\ref{eq:ESF}), so that Corollary~\ref{cor:decomposition} applies. We use a simple rectangular partition $\{\chi_1,\chi_2\}$ of unity, where
\[
\chi_2(t) = \begin{cases}0, & t < a\\ 1, & t\geq a \end{cases}
\]
and $\chi_1(t)=1-\chi_2(t)$. Corollary~\ref{cor:decomposition} gives the corresponding decomposition $S=S_1+S_2$, where $S_i=\ESF(\phi_i,0.5)$. To find the explicit formula for $S_2$, we compute the function $\phi_2$ using Equation~(\ref{eq:phi_j general}). Integrating twice with the choice $u_2=v_2=a$ gives
\begin{align*}
\phi_2(u) 
&= \begin{cases}0, & u<a \\ 2(u-a)^2, & u\geq a  \end{cases} \\
&= 2(u-a)^2\,\one\{u\geq a\}.
\end{align*}
Thus $S_2=\ESF(\phi_2,0.5)$, which yields the explicit formula given by Equation~(\ref{eq:S_2 explicit}) via Equation~(\ref{eq:ESF}). The explicit formula for $S_1$ follows easily from the identity $S_1=S-S_2$.

Figure~\ref{fig:esf decomp} illustrates the decomposition $S=S_1+S_2$ with respect to the rectangular partition $\{\chi_1,\chi_2\}$ of unity with $a=3$, and also with respect to a trapezoidal partition. For each forecast $x$ and observation $y$, the solid line represents the score $S(x,y)$, the dotted line the score $S_1(x,y)$ corresponding to the weight function $\chi_1$ with support on the left of the interval, and the dashed line the score $S_2(x,y)$ corresponding to the weight function $\chi_2$ with support on the right of the interval. Note that when the forecast $x$ and observation $y$ both lie outside the support of $\chi_j$ then $S_j(x,y)=0$.
\end{example}

\begin{figure}[bt]
\centering
\includegraphics{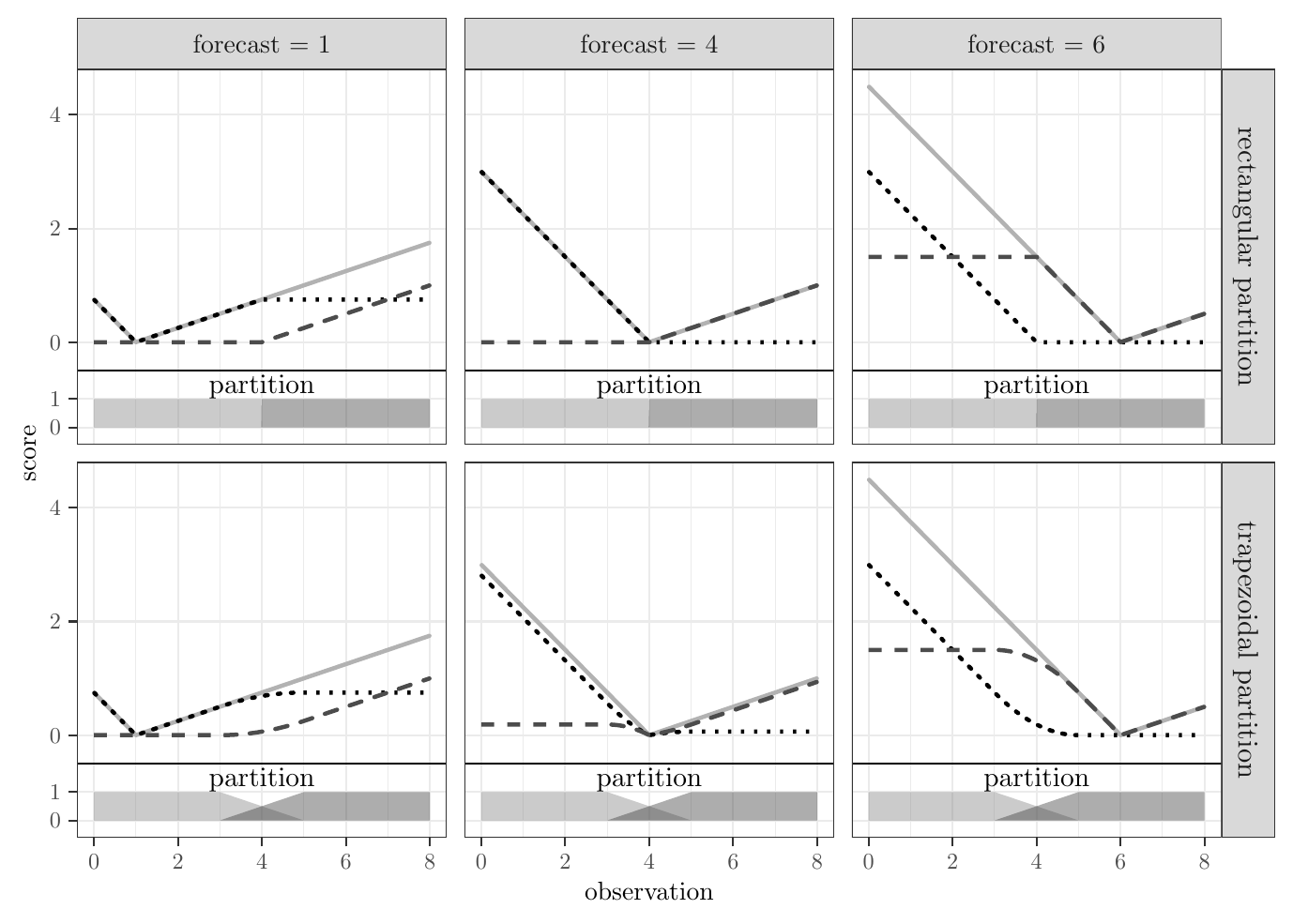}
\caption{Decomposition $S=S_1+S_2$ of the commonly used consistent scoring function for $0.25$-quantile forecasts, using the rectangular and trapezoidal partitions. The solid line represents $S$, the dotted line $S_1$ and the dashed line $S_2$ for different forecast--observation pairs. In each case, $S_1$ emphasises performance on the interval $(0,4)$ while $S_2$ emphasises performance on $(4,8)$.}
\label{fig:qsf decomp}
\end{figure}

\begin{example}[Decomposition of a quantile scoring function]\label{ex:qsf}
Let $S$ denote the scoring function
\[
S(x,y) = 
\begin{cases}
0.25(y-x), & y\geq x\\
0.75(x-y), & x < y,
\end{cases}
\]
so that $S=\QSF(g,\alpha)$ where $g(t)=t$ and $\alpha=0.25$. The decomposition $S=S_1+S_2$ is illustrated in Figure~\ref{fig:qsf decomp} with respect to two different partitions $\{\chi_1,\chi_2\}$ of unity. The solid line represents $S$, the dotted line $S_1$ and the dashed line $S_2$.
\end{example}

\begin{example}[Synthetic data]\label{ex:synthetic}
Suppose that the climatological distribution of an unknown quantity $Y$ is normal with $Y\sim\cN(4,15^2)$. Two forecast systems A and B issue point forecasts for $Y$ targeting the mean functional. Their respective forecast errors are identically independently normally distributed with error characteristics $e_\rA \sim\cN(0, (\arctan(y-10)+2)^2)$ (where $y$ is the observation) and $e_\rB \sim\cN(0, 2^2)$. Hence the standard deviation of $e_\rA$ is lower than the standard deviation of $e_\rB$ when $y$ is less than $10$ while the reverse is true when $y$ is greater than $10$. This is evident in the varying degree of scatter about the diagonal line in forecast--observation plot of Figure~\ref{fig:synthetic}.

We sample 10000 independent observations and corresponding forecasts, and compare both forecast systems using the squared error scoring function $S$ along with the two components of its decomposition $S=S_1+S_2$. The decomposition is induced from a rectangular partition $\{\chi_1,\chi_2\}$ of unity on $\RRR$ with $\supp(\chi_1)=(-\infty,10)$ and $\supp(\chi_2)=[10,\infty)$. The mean scores for each system are shown in Table~\ref{tab:synthetic}, along with a 95\% confidence interval of the difference in mean scores.

An analysis of this sample concludes that neither System A nor System B is significantly better than its competitor as measured by $S$, since $0$ belongs to the confidence interval of the difference in their mean scores. However, one can infer with high confidence that A performs better when scored by $S_1$, since $0$ is well outside the corresponding confidence interval. 
Similarly, one may conclude with high confidence that B performs better when scored by $S_2$.

Based on the results in the table, one would use the forecast from A if both forecasts are less than 10, and the forecast from B if both forecasts are greater than 10. If the forecasts lie on opposite sides of 10, one option is to take the average of the two forecasts. We revisit this example again in Section~\ref{s:murphy} from the perspective of users and optimal decision rules.
\end{example}

\renewcommand{\arraystretch}{1.3}
\begin{table}
	\begin{center}
		\caption{Comparative evaluation of forecast systems A and B from Synthetic data example, rounded to 2 decimal places. The difference is the mean score of A minus the mean score of B.}
		\label{tab:synthetic}
		\begin{tabular}{|l|r|r|r|}
			\hline
			Mean score & System A & System B & 95\% confidence interval of difference \\
			\hline
			$\bar{S}$		& 4.14 & 4.02 & (-0.12, 0.36) \\
			$\bar S_1$	& 0.61 & 2.65 & (-2.16, -1.92) \\
			$\bar S_2$	& 3.53 & 1.36 & (1.97, 2.36) \\
			\hline
		\end{tabular}
	\end{center}
\end{table}

\begin{example}[Sydney rainfall forecasts]\label{ex:sydney all}
Two Bureau of Meteorology forecast systems BoM and OCF issue predictive distributions for daily rainfall at Sydney Observatory Hill. Climatologically, daily rainfall exceeds 35.8\,mm on average 12 times a year, and 42.2\,mm on average 6 times a year. We partition the outcome space $[0,\infty)$ using a trapezoidal partition $\{\chi_1,\chi_2\}$ of unity, where $\chi_2(t)=0$ if $0\leq t<35.8$, $\chi_2(t) = 1$ if $t\geq42.2$ and $\chi_2(t)=(t-35.8)/(42.2-35.8)$ if $35.8\leq t<42.2$. Naturally, $\chi_1(t) = 1 - \chi_2(t)$.

\begin{figure}[bt]
\centering
\includegraphics{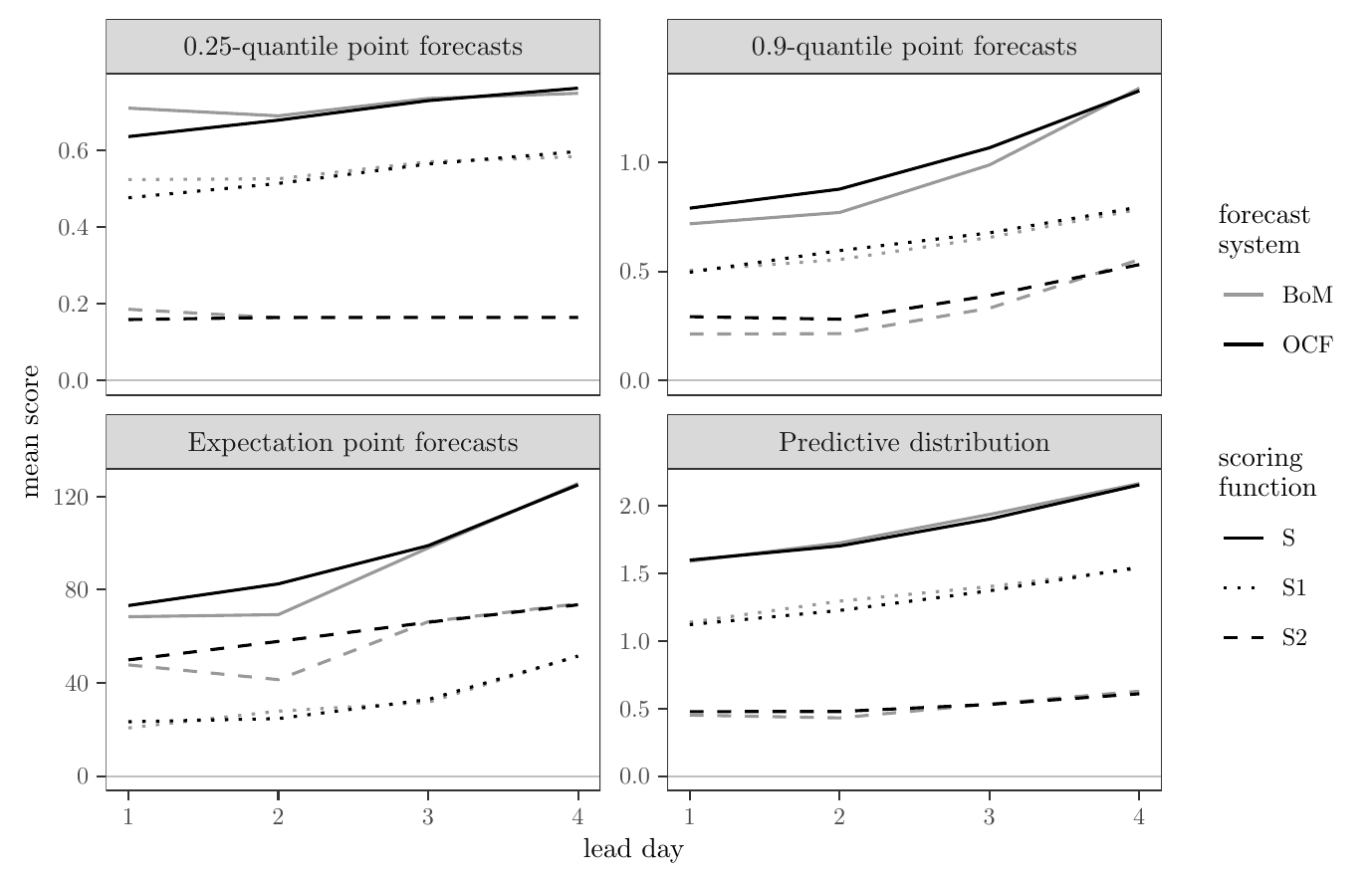}
\caption{Proper score decompositions $S=S_1+S_2$ by lead day for different types of Sydney daily rainfall forecasts, where $S$ is a standard quantile scoring function (top), squared error scoring function (bottom left) or CRPS (bottom right). In each case the score $S_2$ emphasises performance for heavier rainfall.}
\label{fig:precip all}
\end{figure}

Over the entire outcome space, the quality of each system's forecasts is assessed using the CRPS (for predictive distributions), the squared error scoring function (for expectation point forecasts), and the standard quantile scoring function of Equation~(\ref{eq:standard qsf}) (for quantile point forecasts). Moreover, each of these scores $S$ are decomposed as $S=S_1+S_2$ using the common partition $\{\chi_1,\chi_2\}$. Their mean scores by forecast lead day, for the period July 2018 to June 2020, are shown in Figure~\ref{fig:precip all}. This example illustrates that the same partition of unity can be used to hone in on particular regions of interest across a variety of forecast types, using an assessment method that cannot be hedged. When performance is assessed with emphasis on heavier rainfall via $S_2$, BoM is better than OCF at lead days 1 and 2 for expectile and 0.9-quantile forecasts, marginally better with its predictive distributions but worse than OCF at lead day 1 for 0.25-quantile forecasts.
\end{example}

\section{Decision theoretic interpretation of scoring function decompositions}\label{s:murphy}

\textit{Mixture representations} and \textit{elementary scoring functions} facilitate a decision-theoretic interpretation of the scoring function decompositions of Corollary~\ref{cor:decomposition}.

To avoid unimportant technical details, in this section assume that $I=\RRR$, $g$ is a nondecreasing differentiable function on $\RRR$ and $\phi$ is convex twice-differentiable function on $\RRR$. Suppose that $S$ is any one of the scoring functions $\QSF(g,\alpha)$, $\ESF(\phi,\alpha)$ or $\HSF(\phi,\nu)$. Thus $S$ is consistent for some functional $\rT$ (either a specific quantile, expectile or Huber mean). Then $S$ admits a mixture representation
\begin{equation}\label{eq:mixture representation}
S(x,y) = \int_{\RRR} S_{\theta}^{\rT}(x,y)\,\dd\M(\theta),
\end{equation}
where each $S_\theta^\rT$ is an elementary scoring function for $\rT$ and the \textit{mixing measure} $\dd\M(\theta)$ is nonnegative (\citealt{ehm2016quantiles, taggart2021point}). That is, $S$ is a weighted average of elementary scoring functions. Explicit formulae for these elementary scoring functions and mixing measures are given in Tables~\ref{tab:elementary} and \ref{tab:mixing measure}.

The elementary scoring functions $S_\theta^\rT$  and their corresponding functionals $\rT$ arise naturally in the context of optimal decision rules. In a certain class of such rules, a predefined action is taken if and only if the point forecast $x$ for some unknown quantity $Y$ exceeds a certain decision threshold $\theta$. The usefulness of the forecast for the problem at hand can be assessed via a loss function $S_\theta$, where $S_\theta(x,y)$ encodes the economic regret, relative to a perfect forecast, of applying the decision rule with forecast $x$ when the observation $y$ is realised. Typically $S_\theta(x,y)>0$ whenever $\theta$ lies between $x$ and $y$ (e.g., the forecast exceeded the decision threshold but the realisation did not, resulting in regret), and $S_\theta(x,y)=0$ otherwise (i.e., a perfect forecast would have resulted in the same decision, resulting in no regret).

\begin{table}
	\begin{center}
		\caption{Elementary scoring functions $S_\theta^\rT(x,y)$ for different functionals $\rT$.}
		\label{tab:elementary}
		\begin{tabular}{|l|r|}
			\hline
			Functional $\rT$ & Elementary score $S_\theta^\rT(x,y)$ \\
			\hline
			\hline
			$\alpha$-quantile & $\begin{cases}1-\alpha, & y\leq\theta<x \\ \alpha, & x\leq\theta<y \\ 0,&\text{otherwise} \end{cases}$ \\
			\hline
			$\alpha$-expectile & $\begin{cases}(1-\alpha)|y-\theta|,&y\leq\theta<x\\ \alpha|y-\theta|, & x\leq\theta<y \\0,&\text{otherwise}\end{cases}$\\
			\hline
			$\nu$-Huber mean & $\begin{cases}\tfrac12\min(\theta-y,\nu) & y\leq\theta<x \\ \tfrac12\min(y-\theta,\nu), & x\leq\theta<y \\ 0,&\text{otherwise} \end{cases}$ \\
			\hline
		\end{tabular}
	\end{center}
\end{table}

For the decision rule to be optimal over many forecast cases, the point forecast $x$ should be one that minimises the expected score $\EEE S_\theta(x,Y)$, where $Y\sim F$ for the predictive distribution $F$. Binary betting decisions (\citealt{ehm2016quantiles}) and simple cost--loss decision models (where the user must weigh up the cost of taking protective action as insurance against an event which may or may not occur (\citealt{richardson2000skill})) give rise to some $\alpha$-quantile of $F$ being the optimal point forecast. The $\alpha$-expectile functional gives the optimal point forecast for investment problems with cost basis $\theta$, revenue $y$ and simple taxation structures (\citealt{ehm2016quantiles}). The Huber mean generates optimal point forecasts when profits and losses are capped in such investment problems (\citealt{taggart2021point}). Though the language here is financial, such investment decisions can be made using weather forecasts (\citealt[Example 5.4]{taggart2021point}). In each case, the particular score $S_\theta(x,y)$ is, up to a multiplicative constant, the elementary score $S_\theta^\rT(x,y)$ in Equation (\ref{eq:mixture representation}) for the appropriate functional $\rT$.

Thus the representation (\ref{eq:mixture representation}) expresses $S(x,y)$ as a weighted average of elementary scores, each of which encodes the economic regret associated with a decision made using the forecast $x$ with decision threshold $\theta$. The weight on each $\theta$ is determined by the mixing measure $\dd\M(\theta)$, as detailed in Table~\ref{tab:mixing measure}. Relative to $\QSF(g,\alpha)$, the scoring function $\QSF(g_j,\alpha)$ weighs the economic regret for decision threshold $\theta$ by a factor of $\chi_j(\theta)$, thus privileging certain decision thresholds over others.

\begin{table}
	\begin{center}
		\caption{Weights in the mixing measure $\dd\M(\theta)$ for different scoring functions $S$.}
		\label{tab:mixing measure}
		\begin{tabular}{|l|r|}
			\hline
			Scoring function $S$ & $\dd\M(\theta)$ \\
			\hline
			\hline
			$\QSF(g,\alpha)$ & $g'(\theta)\,\dd\theta$ \\
			$\QSF(g_j,\alpha)$ & $\chi_j(\theta)g'(\theta)\,\dd\theta$ \\
			\hline
			$\ESF(\phi,\alpha)$, $\HSF(\phi,\alpha)$ & $\phi''(\theta)\,\dd\theta$ \\
			$\ESF(\phi_j,\alpha)$, $\HSF(\phi_j,\alpha)$ & $\chi_j(\theta)\phi''(\theta)\,\dd\theta$ \\
			\hline
		\end{tabular}
	\end{center}
\end{table}

\begin{figure}[bt]
\centering
\includegraphics{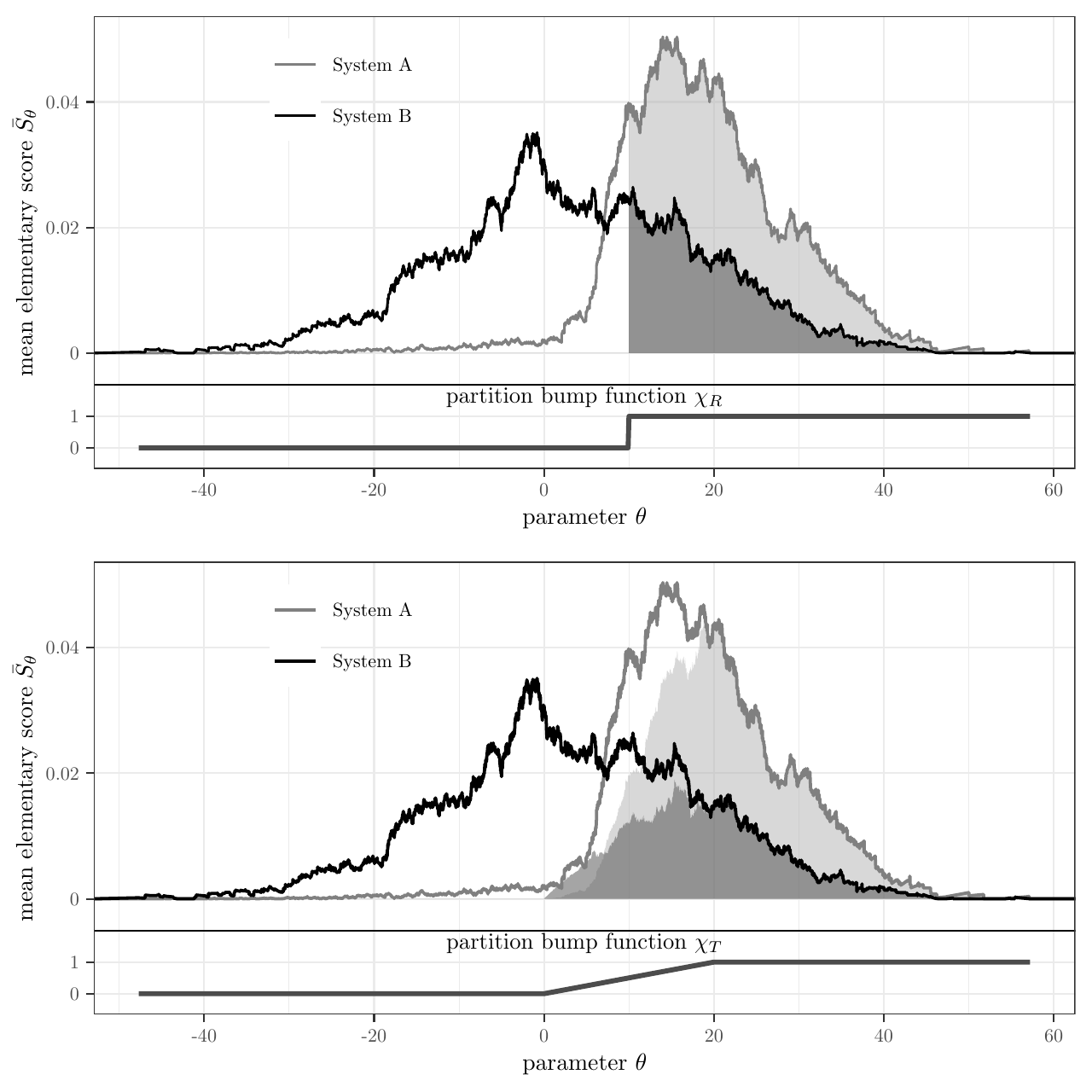}
\caption{Murphy diagrams for the two forecast systems A (lighter) and B (darker) of the Synthetic data example. The graphs on each panel are the same but the shaded regions differ. Each shaded area is proportional to the mean score of the scoring function $\ESF_j(t\mapsto 2t^2, 1/2)$, which is the summand in the mean squared error decomposition corresponding to the weight function $\chi$. Here $\chi$ is either the rectangular weight function $\chi_R$ (top) or trapezoidal weight function $\chi_T$ (bottom).}
\label{fig:murphy}
\end{figure}

We illustrate these ideas using the point forecasts generated by Systems A and B from the Synthetic data example (Section~\ref{ss:examples}).  These point forecasts target the mean functional, and so induce optimal decision rules for investment problems with cost basis $\theta$. The decision rule is to invest if and only if the forecast future value $x$ of the investment exceeds the fixed up-front cost $\theta$ of the investment. 
As previously noted, neither forecast system performs significantly better than the other when scored by squared error. However, for certain decision thresholds $\theta$, the mean elementary score $\bar S_\theta$ (which measures the average economic regret) of one system is clearly superior to the other. This is illustrated by the \textit{Murphy diagram} of Figure~\ref{fig:murphy}, which is a plot of $\bar S_\theta$ against $\theta$ (\citealt{ehm2016quantiles}). Since a lower score is better, users with a decision threshold exceeding about $8$ should use forecasts from B, while those with a decision threshold less than $8$ should use forecasts from A.

Let $S=S_1+S_2$ denote the same decomposition of the squared error scoring function used for the Synthetic data example (Section~\ref{ss:examples}), induced from a rectangular partition of unity. To aid simplicity of interpretation, assume for the user group under consideration that there is one constant $k$ such that the each elementary score equals $k$ multiplied by the economic regret. We observe the following.

For $S$, $\phi(t)=2t^2$ and so $\dd\M(\theta) = 4\,\dd\theta$ by Table~\ref{tab:mixing measure}. So by Equation~(\ref{eq:mixture representation}), the mean squared error $\bar S$ for each forecast system is proportional to the total area under the respective Murphy graph. Hence mean squared error $\bar S$ can be interpreted as being proportional to the mean economic regret averaged across all decision thresholds.

For $S_2$, $\dd\M(\theta)=4\chi_2(\theta)\,\dd\theta$, where $\chi_2$ is the weight function $\chi_R$ on the top panel of Figure~\ref{fig:murphy}. From Equation~(\ref{eq:mixture representation}), the mean score $\bar{S}_2$ for each forecast system is proportional to the area under the respective Murphy curve for decision thresholds $\theta$ in $[10,\infty)$. This is illustrated on the top panel of Figure~\ref{fig:murphy} with shaded regions. System B has the smaller area and mean score and is hence superior. So $\bar{S}_2$ can be interpreted as being proportional to the mean economic regret averaged across all decision thresholds $\theta$ in $[10,\infty)$. Similarly, $\bar{S}_1$ can be interpreted as being proportional to the mean economic regret averaged across all decision thresholds $\theta$ less than $10$.

The bottom panel of Figure~\ref{fig:murphy} illustrates the effect of using a trapezoidal weight function $\chi_T$. The corresponding scoring function applies zero weight to economic regret experienced by users with decision thresholds less than 0, and increasing weight to thresholds greater than zero until a full weight of 1 is applied for decision thresholds greater than or equal to 20. The areas of the shaded regions are proportional to the mean score for Systems A and B.

\section{Conclusion}

We have shown that the scoring functions that are consistent for quantiles, expectations, expectiles or Huber means can be decomposed as a sum of consistent scoring functions, each of which emphasises predictive performance in a region of the variable's range according to the selected partition of unity. In particular, this allows the comparison of predictive performance for the extreme region of a variable's range in a way that cannot be hedged and is not susceptible to misguided inferences.

The decomposition is consonant with the analogous decomposition of the CRPS using the threshold-weighting method, and hence by extension with known decompositions for the absolute error scoring function.

Such decompositions are a consequence of the mixture representation for each of the aforementioned classes of consistent scoring functions and their associated functionals. This has two implications. First, each score in the decomposition can be interpreted as the weighted average of economic regret associated with optimal decision rules, where the average is taken over all user decision thresholds and the weight applied is given by the corresponding weight function in the partition of unity. Second, such decompositions will also exist for the consistent scoring functions of other functionals that have analogous mixture representations.

\section*{Acknowledgements}
The author thanks Jonas Brehmer, Deryn Griffiths, Michael Foley and Tom Pagano for their constructive comments, which helped improve the quality of this paper.

\appendix

\section{Proof of Corollary 1}\label{s:proof}

As mentioned in Section~\ref{ss:decomposition of sf}, Corollary~\ref{cor:decomposition} essentially follows from the mixture representation of theorems of \cite{ehm2016quantiles} and \cite{taggart2021point} by decomposing the mixing measure. However, we provide a direct proof to avoid some technical hypotheses of those theorems.

\begin{proof}
We prove for the case $\ESF$. The proof for the cases $\QSF$ and $\HSF$ are similar. 

To begin, note that if $\phi(s)=\psi(s)+cs +d$ whenever $s\in I$ for some real constants $c$ and $d$ then $\ESF(\phi,\alpha)=\ESF(\psi,\alpha)$. So without loss of generality we may assume that $u_j=u_0$ and $v_j=v_0$ in the definition of $\phi_j$ whenever $1\leq j\leq n$ for some constants $u_0$ and $v_0$ in $I$.

Suppose that $\phi$ is a convex twice-differentiable function and $\alpha\in(0,1)$. Given the partition of unity $\{\chi_j\}_{j=1}^n$, define $\phi_j$ by Equation (\ref{eq:phi_j general}). Then
\begin{align*}
&|\one\{y<x\} - \alpha|^{-1}\sum_{j=1}^n \ESF(\phi_j,\alpha)(x,y) \\
&\qquad=  \sum_{j=1}^n \big(\phi_j(y) - \phi_j(x) - \phi_j'(x)(y-x)\big) \\
&\qquad= \sum_{j=1}^n \left( \int^y_x\int^v_{v_0}\chi_j(\theta)\phi''(\theta)\,\dd\theta\,\dd v - (y-x)\int^x_{v_0} \chi_j(\theta)\phi''(\theta)\,\dd\theta\right) \\
&\qquad= \int^y_x\int^v_{v_0}\phi''(\theta)\,\dd\theta\,\dd v - (y-x)\int^x_{v_0}\phi''(\theta)\,\dd\theta \\
&\qquad= \int^y_x \big(\phi'(v)-\phi'(v_0)\big)\,\dd v - (y-x)\big(\phi'(x)-\phi'(v_0)\big) \\
&\qquad= (\phi(y)-\phi'(v_0)y) - (\phi(x)-\phi'(v_0)x) - (y-x)(\phi'(x)-\phi'(v_0)) \\
&\qquad= |\one\{y<x\} - \alpha|^{-1} \ESF(\phi,\alpha)(x,y),
\end{align*}
which establishes Equation (\ref{eq: ESF decomposition}). The convexity of $\phi_j$ follows from the fact that $\phi_j''(u)=\chi_j(u)\phi''(u)\geq0$, since $\phi$ is convex. Finally, suppose that $x,y\in I_0\subset I$ and $\supp(\chi_j) \cap I_0 = \emptyset$. Now
\begin{align*}
|\one\{y<x\} - \alpha|^{-1}\ESF(\phi_j,\alpha)(x,y)
&= \int_x^y (\phi_j'(w) - \phi_j'(x))\,\dd w \\
&= \int_x^y \int_x^w \phi_j''(z) \,\dd z\,\dd w \\
&= \int_x^y \int_x^w \chi_j(z)\phi''(z) \,\dd z\,\dd w,
\end{align*}
and since $w$ lies between $x$ and $y$, it follows that $w$ also lies in $I_0$. But $\chi_j=0$ on $I_0$ and hence the inner integral vanishes. This shows that $\ESF(\phi_j,\alpha)=0$ on $I_0\times I_0$.
\end{proof}

\section{Formulae for commonly used scoring functions} \label{s:g_j and phi_j}

Table~\ref{tab:closed-form} presents closed-form expressions for $g_j$ of Equation~(\ref{eq:g_j}) in the case when $g(t)=t$ and for $\phi_j$ of Equation~(\ref{eq:phi_j general}) in the case when $\phi(t)=2t^2$, each induced by rectangular weight functions of the form  (\ref{eq:rectangular bump}) or trapezoidal weight functions of the form  (\ref{eq:trapezoidal bump}). These expressions facilitate the computation of decompositions for the absolute error, standard quantile, squared error and Huber loss  scoring functions. Note also that for each weight function we have $\phi_j'(t)=4g_j(t)$.

\begin{table}
	\begin{center}
	\caption{Closed-form expressions for $g_j$ and $\phi_j$ for different weight functions when $g(t)=t$, $\phi(t)=2t^2$.}
		\label{tab:closed-form}
		\begin{footnotesize}
		\begin{tabular}{|l|r|}
		\hline
		weight function & expression \\ \hline \hline
		\multirow{2}{*}{$\mat{rectangular,}{Equation~(\ref{eq:rectangular bump})}$} & $\gjrect$ \\ \cline{2-2} 
		                  &  $\phijrect$ \\ \hline
		\multirow{2}{*}{$\mat{trapezoidal,}{Equation~(\ref{eq:trapezoidal bump})}$} & $\gjtrap$ \\ \cline{2-2} 
		                  &  $\phijtrap$ \\ \hline
		\end{tabular}
		\end{footnotesize}
	\end{center}
\end{table}



\begin{thebibliography}{}

\bibitem[Bellini and Di~Bernardino, 2017]{bellini2017risk}
Bellini, F. and Di~Bernardino, E. (2017).
\newblock Risk management with expectiles.
\newblock {\em The European Journal of Finance}, 23(6):487--506.

\bibitem[Ehm et~al., 2016]{ehm2016quantiles}
Ehm, W., Gneiting, T., Jordan, A., and Kr{\"u}ger, F. (2016).
\newblock Of quantiles and expectiles: consistent scoring functions, choquet
  representations and forecast rankings.
\newblock {\em J. R. Statist. Soc. B}, 78:505--562.

\bibitem[Gneiting, 2011a]{gneiting2011making}
Gneiting, T. (2011a).
\newblock Making and evaluating point forecasts.
\newblock {\em Journal of the American Statistical Association},
  106(494):746--762.

\bibitem[Gneiting, 2011b]{gneiting2011quantiles}
Gneiting, T. (2011b).
\newblock Quantiles as optimal point forecasts.
\newblock {\em International Journal of forecasting}, 27(2):197--207.

\bibitem[Gneiting and Katzfuss, 2014]{gneiting2014probabilistic}
Gneiting, T. and Katzfuss, M. (2014).
\newblock Probabilistic forecasting.
\newblock {\em Annual Review of Statistics and Its Application}, 1:125--151.

\bibitem[Gneiting and Ranjan, 2011]{gneiting2011comparing}
Gneiting, T. and Ranjan, R. (2011).
\newblock Comparing density forecasts using threshold-and quantile-weighted
  scoring rules.
\newblock {\em Journal of Business \& Economic Statistics}, 29(3):411--422.

\bibitem[Huber, 1964]{huber1964robust}
Huber, P.~J. (1964).
\newblock Robust estimation of a location parameter.
\newblock {\em Annals of Mathematical Statistics}, 35:73--101.

\bibitem[Lerch et~al., 2017]{lerch2017forecaster}
Lerch, S., Thorarinsdottir, T.~L., Ravazzolo, F., and Gneiting, T. (2017).
\newblock Forecaster’s dilemma: Extreme events and forecast evaluation.
\newblock {\em Statistical Science}, 32(1):106--127.

\bibitem[Newey and Powell, 1987]{newey1987asymmetric}
Newey, W.~K. and Powell, J.~L. (1987).
\newblock Asymmetric least squares estimation and testing.
\newblock {\em Econometrica}, 55:819--847.

\bibitem[Patton, 2020]{patton2020comparing}
Patton, A.~J. (2020).
\newblock Comparing possibly misspecified forecasts.
\newblock {\em Journal of Business \& Economic Statistics}, 38(4):796--809.

\bibitem[Richardson, 2000]{richardson2000skill}
Richardson, D.~S. (2000).
\newblock Skill and relative economic value of the {ECMWF} ensemble prediction
  system.
\newblock {\em Quarterly Journal of the Royal Meteorological Society},
  126(563):649--667.

\bibitem[Savage, 1971]{savage1971elicitation}
Savage, L.~J. (1971).
\newblock Elicitation of personal probabilities and expectations.
\newblock {\em Journal of the American Statistical Association},
  66(336):783--801.

\bibitem[Sharpe et~al., 2020]{sharpe2020new}
Sharpe, M., Bysouth, C., and Gill, P. (2020).
\newblock New operational measure to assess extreme events using site-specific
  climatology.
\newblock Presented at 2020 International Verification Methods Workshop Online.
  Retreived from https://jwgfvr.univie.ac.at/presentations-and-notes/ on 5
  January 2021.

\bibitem[Taggart, 2021]{taggart2021point}
Taggart, R. (2021).
\newblock Point forecasting and forecast evaluation with generalized {H}uber
  loss.
\newblock {\em Electronic Journal of Statistics}, to appear.

\bibitem[Thomson, 1979]{thomson1979eliciting}
Thomson, W. (1979).
\newblock Eliciting production possibilities from a well-informed manager.
\newblock {\em Journal of Economic Theory}, 20:360--380.

\end{thebibliography}

\end{document}